\documentclass[conference]{IEEEtran}
\IEEEoverridecommandlockouts
\usepackage{cite}
\usepackage{amsmath,amssymb,amsfonts}
\usepackage{algorithmic}
\usepackage{graphicx}
\usepackage{textcomp}
\usepackage{xcolor}
\usepackage{subcaption}
\def\BibTeX{{\rm B\kern-.05em{\sc i\kern-.025em b}\kern-.08em
    T\kern-.1667em\lower.7ex\hbox{E}\kern-.125emX}}
\begin{document}

\title{A SER-based Device Selection Mechanism in Multi-bits Quantization Federated Learning\\
}


\author{Pengcheng Sun, 
Erwu Liu, \IEEEmembership{Senior Member,~IEEE,}
Rui Wang, \IEEEmembership{Senior Member,~IEEE,}
\thanks{Corresponding author: Erwu Liu.
}
}

\maketitle

\begin{abstract}
The quality of wireless communication will directly affect the performance of federated learning (FL), so this paper analyze the influence of wireless communication on FL through symbol error rate (SER). In FL system, non-orthogonal multiple access (NOMA) can be used as the basic communication framework to reduce the communication congestion and interference caused by multiple users, which takes advantage of the superposition characteristics of wireless channels. The Minimum Mean Square Error (MMSE) based serial interference cancellation (SIC) technology is used to recover the gradient of each terminal node one by one at the receiving end. In this paper, the gradient parameters are quantized into multiple bits to retain more gradient information to the maximum extent and to improve the tolerance of transmission errors. On this basis, we designed the SER-based device selection mechanism (SER-DSM) to ensure that the learning performance is not affected by users with bad communication conditions, while accommodating as many users as possible to participate in the learning process, which is inclusive to a certain extent. The experiments show the influence of multi-bit quantization of gradient on FL and the necessity and superiority of the proposed SER-based device selection mechanism.
\end{abstract}


\section{Introduction}

The centralized Federated Learning (FL) architecture allows distributed devices to compute local model updates from local data sets and then achieve convergence of learning network by  interacting with a parameter server (base station (BS) as the parameter server in this article) iteratively \cite{FLfoundamental}. However, due to the limitation of communication resources, wireless communication conditions become one of the decisive factors of quality of FL network \cite{ComFL}. Non-orthogonal multiple access (NOMA) technology can effectively alleviate the communication congestion caused by multiple users in FL by means of multi-user parallel transmission \cite{FLuser}. In addition, the serial interference cancellation based on minimum mean square error (MMSE-SIC) in this paper can recover the local gradient parameters uploaded by each user while eliminating inter-user interference.

In this paper, the symbol error rate (SER) \cite{SER} of each device in gradient transmission is used as the bridge between wireless communication and FL network, which reflects the influence of communication factors (such as quantization and modulation mode) on FL performance. This paper proposes an inclusive SER-based device selection mechanism (SER-DSM), which alleviates the straggler problem caused by the heterogeneity of user communication conditions in FL \cite{yuan}. The device selection mechanism proposed by \cite{TWC} based on packet error rate \cite{packet}, that is, the parameter server will reject the data uploaded by the user who makes any error in gradient parameters, has improvement on FL performance to some extent compared to not selecting the user, but it may still produce data waste and lack inclusiveness. The selection mechanism proposed in this paper allows more users with less than acceptable SER to participate in FL updates, which not only avoids data waste, but also gets better FL performance. The FL convergence based on SER-DSM is further analyzed, and the feasibility is verified theoretically.

The basis of the inclusive device selection mechanism proposed in this paper is that the user device as the sender performs multi-bit quantization of the local model parameters. Compared with the one-bit scheme \cite{onebit}, the multi-bit quantization scheme has a higher fault tolerance rate and can make the transmitted symbol contain more information about the parameter gradient, which is conducive to the learning network. Experiments analysis the effects of different order of quantization/modulation schemes on FL performance firstly, and then the necessity and superiority of SER-DSM proposed in this paper are demonstrated.

\section{System Model}

We consider a FL system consisting of \textit{K} intelligent devices and one BS equipped with \textit{N} antennas. The number of devices selected to participate in the model training is $\mathcal{K}$ of \textit{K} $(1\le\mathcal{K}\le K)$. The set of selected devices is indexed by $k=\left\{1,2,\ldots,\ \mathcal{K}\right\}$. The $k$-th device has its own local data set of size $\mathcal{D}_k$, consisting of labeled data samples $\left\{\left(\boldsymbol{x}_k,y_k\right)\right\}\in\mathcal{D}_k$, where $\boldsymbol{x}_k\in\mathbb{R}^d$ denotes the input unlabeled data vector of the FL algorithm and $y_{ks}\in\mathbb{R}$ the associated label (output of $\boldsymbol{x}_k$). The size of $\boldsymbol{x}_k$ depends on the specific FL task. For simplicity, we consider an FL algorithm with a single output. A vector $\boldsymbol{w}_k$ is defined to capture the parameters related to the local FL model that is trained by $\boldsymbol{x}_k$ and $y_k$. 

\subsection{FL System}

This paper uses the traditional FL update process. The local model parameter of the $k$-th user in the $n$-th communication round can be updated via gradient descent based on
\begin{equation}
\boldsymbol{w}_k^{\left[n+1\right]}=\boldsymbol{w}_k^{\left[n\right]}-\lambda\cdot\boldsymbol{g}_k^{\left[n\right]},
\label{w_globe}
\end{equation}
where $\lambda$ denotes the learning rate and $\textbf{g}$ is gradient of local sample loss. Then FL aggregates the local gradient parameters via FedAvg \cite{FedAvg}. If the local model parameter can be transmitted reliably to the BS, the global model parameter would be calculated with relative precision as
\begin{equation}
    \boldsymbol{w}^{\left[n+1\right]}=\frac{1}{\mathcal{D}}\sum_{k=1}^{K}\mathcal{D}_k\boldsymbol{w}_k^{\left[n+1\right]}
.\end{equation}

The process in the $n$-th communication round keeps iterating until the convergence condition is satisfied. Since all of the local FL models are transmitted over wireless communication, they may contain erroneous symbols due to the unreliable nature of the wireless channel, which will have a significant impact on the performance of FL. 

\subsection{Communication System based on NOMA}

\textbf{Transmission process}: We assume a block fading channel model, where the channel coefficients remain invariant during the whole FL training process. Let $\boldsymbol{h}_{k}\in\mathbb{C}^N$ denote the direct $k$-th-device to BS channel coefficient vector, and the amplitudes are assumed to be independent Rician random variables. We assume perfect channel state information (CSI) at the BS and devices. 

In digital communication, each device needs to use a limited number of digital symbols to transmit the local gradients, which requires each device to quantify each local gradient parameter symbol into a sequence of 1 bit or multiple bits of digital symbols at each training round. For the 1-bit quantization scheme \cite{huangkaibin}, although the transmission efficiency can be greatly improved, the receiver can not accurately recover the original local gradient parameters, which is not desirable in some scenarios requiring high precision. In this paper, the local gradient parameters are quantized into multiple bits, so that the receiver can accurately recover the original local gradient parameters from each user. In addition, this scheme will also effectively improve the fault tolerance of transmission compared to that of 1-bit quantization. Specifically, if the transmission error occurs in the low region of the multi-bit string, the gradient parameters recovered by the receiver are not much different from the real values. When the aggregation method with the average character is used in FL, this deviation will be further reduced, which will provide the possibility for FL to accommodate more participants. In fact, the device selection mechanism proposed in Section III of this paper is based on this.

In the model aggregation of the $n$-th training round, we denote the coded signal sequences from the active devices by $\boldsymbol{x}_k^{\left[n\right]}=\left[x_{k,1}^{\left[n\right]},x_{k,2}^{\left[n\right]},\ldots,x_{k,q}^{\left[n\right]}\right]^{T}, k\in\mathcal{K}$, where $q$ is the size of parameter. We note that, $x_{k,i}^{\left[n\right]}$ is a code set of $b_{signal}$ bits. Each of the $b_{signal}$-bits-code sequence is modulated into a $M$-ary QAM symbol element (where $M=2^{b_{mod}}$, $b_{signal}=\alpha\cdot b_{mod}$, $\alpha>0$ is a positive integer), which is denoted as $s_{k,i}^{\left[n\right]}$. Thus, the channel input vector of the k-th device is $\boldsymbol{s}_k^{\left[n\right]}=\left[s_{k,1}^{\left[n\right]},s_{k,2}^{\left[n\right]},\ldots,s_{k,q}^{\left[n\right]}\right]^{T}$.

\textbf{Receiving mechanism}: The received signal in time slot $d$  at the BS is denoted by
\begin{equation}
    \boldsymbol{y}^{\left[n\right]}\left[d\right]=\sum_{k=1}^{K}{\boldsymbol{h}_{k} p_k s_k^{\left[n\right]}\left[d\right]}+\boldsymbol{n}_0^{\left[n\right]}\left[d\right]
,\end{equation}
where $p_k\in\mathbb{C}$ is the transmitter scalar  of the $k$-th-device, and $\boldsymbol{n}_0^{\left[n\right]}\left[d\right]\in\mathbb{C}^{N\times1}$ is the additive white Gaussian noise (AWGN) vector of the channel with the entries following the distribution of $\mathcal{CN}\left(0,\sigma_n^2\right)$. 

\begin{figure}[htbp]
\centerline{\includegraphics[width=1\linewidth]{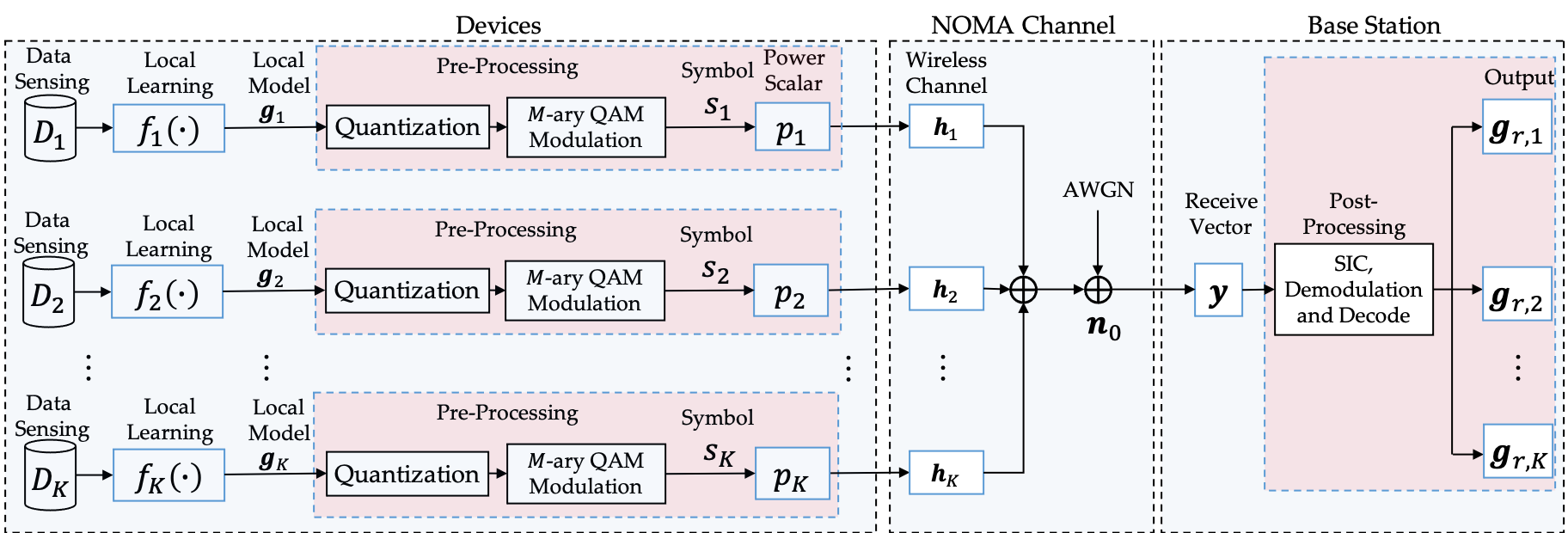}}
\caption{Federated learning under the NOMA computation framework.}
\label{NOMA framework}
\end{figure}

Since all the devices are served by entire bandwidth simultaneously, the user detection and interference management are essential issues for the hybrid signals. In this paper, MMSE-SIC is used in the receiver to perform the detection of multi-user mixed signals. With the help of the SIC technique, the BS can first demodulate the signals of relatively strongest user in a successive order to remove the interference, and then it is able to obtain the superposed signal of weak devices. By doing so, BS could separate the mixed signals and eliminate interference before performing global model aggregation. The block diagram of communication system in this paper is illustrated in Fig.~\ref{NOMA framework}. We assume symbol-level synchronization among the transmitted devices through a synchronization channel. 

\section{Device Selection Mechanism and Analysis of  Convergence based on SER}

\subsection{Device Selection Mechanism}

In this paper, we discuss the average symbol error rate (SER) of the gradients of the $k$-th device through multi-path fading channel with AWGN and MAI to characterize the impact of NOMA communication process on model aggregation performance. The SER of the\ $k$-th device can be expressed as
\begin{equation}
    {SER}_k=1-\left(1-P_{\sqrt M}\right)^2
\label{SER_k}
,\end{equation}
where $P_{\sqrt M}=2\left(1-\frac{1}{\sqrt M}\right)Q\left[\frac{3\mathcal{E}_{av,k}}{\left(M-1\right)\sigma_{n}^2}\right]$ and $\mathcal{E}_{av,k} / \sigma_{n}^2$ denotes the received symbol SINR. In the process of signal detection (equalization) at the receiving end, the linear reception vector of order k is obtained by applying SIC and linear Minimum Mean Square Error (MMSE) filter, i.e., 
\begin{equation}
\begin{split}
\boldsymbol{r}_k={p_k\boldsymbol{h}_k^H\left(\sum_{s=1}^{K}{p_s^2\boldsymbol{h}_s\boldsymbol{h}_s^H}+\sigma_n^2I\right)}^{-1}
\end{split}
.\end{equation}
In this paper, we choose to use the ranking based on channel gain to detect the received signal, where the signal interference to noise ratio of the output of the $k$-th symbol can be expressed as
\begin{equation}
    {SINR}_k=\frac{p_{k}^2\left|\boldsymbol{r}_{k}\boldsymbol{h}_{k}\right|^2}{\sum_{i=k+1}^K\boldsymbol{p}_{i}^2\left|\boldsymbol{r}_{k}\boldsymbol{h}_{i}\right|^2+\|\boldsymbol{r}_{k}\|^2 \sigma_n^2}.
\label{SINRk}
\end{equation}
Thus, substituting \eqref{SINRk} into \eqref{SER_k}, we can obtain the specific expression of $SER_k$ as shown in \eqref{SER_k2}, where $0\leq{SER}_k\leq1,\ \forall k\in K$.
\begin{figure*}[t]
        \begin{equation}
        {SER}_k=1-\left(1-2\left(1-\frac{1}{\sqrt{M}}\right)Q\left(\sqrt{\frac{3}{M-1}\cdot\frac{p_{k}^2\left|\boldsymbol{r}_{k}\boldsymbol{h}_{k}\right|^2}{\sum_{i=k+1}^K\boldsymbol{p}_{i}^2\left|\boldsymbol{r}_{k}\boldsymbol{h}_{i}\right|^2+\|\boldsymbol{r}_{k}\|^2 \sigma_n^2}}\right)\right)^2
    \label{SER_k2}
    \end{equation}
\end{figure*}

If the error symbols generated by transmission in $\boldsymbol{w}_k$ received by BS are beyond our acceptable range (i.e. ${tr}_{SER}\le{SER}_k\le1$), set $C\left(\boldsymbol{w}_k\right)=0$, otherwise set $C\left(\boldsymbol{w}_k\right)=1$, where $C\left(\boldsymbol{w}_k\right)$ describes the degree of data error on BS for the local model transmitted by the $k$-th device. In the actual system, when $C\left(\boldsymbol{w}_k\right)=0$, BS will not use them to update the global model, and will not require the corresponding user to re-transmit the local model data, but only update with the remaining parameters. The advantage of this method is that under the premise of ensuring communication efficiency, more users can be accommodated for global model update to a certain extent. This is in contrast to the case where BS abandons the user's data if any error occurs in that user's transmitted parameter. Thus, taking into account the symbol transmission error, the global model can be written as 
\begin{equation}
\boldsymbol{w}=\frac{\sum_{k=1}^{K}{\mathcal{D}_ka_k\boldsymbol{w}_kC\left(\boldsymbol{w}_k\right)}}{\sum_{k=1}^{K}{\mathcal{D}_ka_kC\left(\boldsymbol{w}_k\right)}}
,\end{equation}
where
\begin{equation}
    C\left(\boldsymbol{w}_k\right) = 
\begin{cases}
  1, & 0\le{SER}_k\le{tr}_{SER} \\
  0, & {tr}_{SER}\le{SER}_k\le1
\end{cases}
,
\end{equation}
$\boldsymbol{a}=\left[a_1,a_2,\ldots,a_K\right]$ is the vector of the user selection index with $a_k=1$ indicating that the $k$-th device performs the FL algorithm and $a_k=$0 otherwise. $\sum_{k=1}^{K}{\mathcal{D}_ka_kC\left(\boldsymbol{w}_k\right)}$ is the total number of training data samples, which depends on the device selection vector $\boldsymbol{a}$ and symbol transmission $C\left(\boldsymbol{w}_k\right)$.

\subsection{Analysis of FL Convergence}

From the above analysis, we know that the performance of FL algorithm is affected by many aspects, such as device selection and communication conditions, and we show their influence through convergence analysis.

Some assumptions for the analysis of FL convergence are made as follows: (1) the loss function $F\left(\boldsymbol{w}\right)$ is a strongly convex function of the parameter $\mu >0$; (2) the gradient function $\nabla F\left(\boldsymbol{w}\right)$ is uniformly Lipschitz continuous for the model parameter $\boldsymbol{w}$; (3) $F\left(\boldsymbol{w}\right)$ is second order continuously differentiable; and (4) $\|\nabla f\left(\boldsymbol{x}_s,y_s; \boldsymbol{w}^{\left[n\right]}\right)\|^2 \leq \zeta_1+\zeta_2\|\nabla F\left(\boldsymbol{w}^{\left[n\right]}\right)\|^2$, $\zeta_1, \zeta_2 \geq 0$. In this paper, the cross entropy function is used as the loss function, which satisfies the above assumptions. After mathematical derivation, a theorem about the FL convergence can be obtained as:

\textbf{Theorem}: given the transmission power $\boldsymbol{p}$, the device selection $\boldsymbol{a}$, the optimal global FL model parameter $\boldsymbol{w}^\ast$ (that is, the global FL model obtained assuming with no transmission errors), and the learning rate $\lambda=1/L$ ($L>0$ is the Lipschitz continuity), there is an upper bound of $\mathbb{E}\left[F\left(\boldsymbol{w}^{\left[n+1\right]}\right)-F\left(\boldsymbol{w}^\ast\right)\right]$, i.e. \eqref{theorem}. Where $\Xi_k=\sum_{m=0}^{q\cdot{tr}_{SER}}{C_{q}^{m}\cdot\left({SER}_k\right)^m\cdot\left(1-{SER}_k\right)^{q-m}}$, $A=1-\frac{\mu}{L}+\frac{4\mu\zeta_2}{LD}\sum_{k=1}^{K}{\mathcal{D}_k\cdot\left(1-a_k\cdot\Xi_k\right)}$, $D=\sum_{k=1}^{K}\mathcal{D}_k$ is the total size of all user training data, $N$ is the total number of iterations, and $\mathbb{E}\left[\cdot\right]$ is to get expectation.

\begin{figure*}
        \begin{equation}
    \begin{split}
    \mathbb{E}\left[F\left(\boldsymbol{w}^{\left[n+1\right]}\right)-F\left(\boldsymbol{w}^\ast\right)\right] &\leq \frac{2\zeta_1}{L\mathcal{D}}\sum_{k=1}^{K}{\mathcal{D}_k\left(1-a_k\cdot\Xi_k\right)}\frac{1-A^N}{1-A}+A^N\mathbb{E}\left[F\left(\boldsymbol{w}^{\left[0\right]}\right)-F\left(\boldsymbol{w}^\ast\right)\right]
    \end{split}
    \label{theorem}
    \end{equation}
\end{figure*}

From the theorem, we know that the upper bound of $\mathbb{E}\left[F\left(\boldsymbol{w}^{\left[n+1\right]}\right)-F\left(\boldsymbol{w}^\ast\right)\right]$ is mainly related to SER and the device selection $\textbf{a}_k$. In addition, in order to ensure the convergence of FL, we need to set $A<1$. At this point, $\lim\limits_{N\to\infty}{A^N}=0$ can be obtained. From assumption (3), we can get $\mu<L$. So when
\begin{equation}
    \frac{4\mu\zeta_2}{LD}\sum_{k=1}^{K}{\mathcal{D}_k\cdot\left(1-a_k\cdot\Xi_k\right)}<\frac{\mu}{L}
,\end{equation}
FL guarantees convergence. From the above analysis, we can know that the upper bound of $\mathbb{E}\left[F\left(\boldsymbol{w}^{\left[n+1\right]}\right)-F\left(\boldsymbol{w}^\ast\right)\right]$ converges with rate $A$ when $A<1$. In other word, $A$ describes the rate of convergence of the FL algorithm.

\section{Simulation Results}

\subsection{Simulation Setup}

\begin{figure}[htbp]
\centerline{\includegraphics[width=0.6\linewidth]{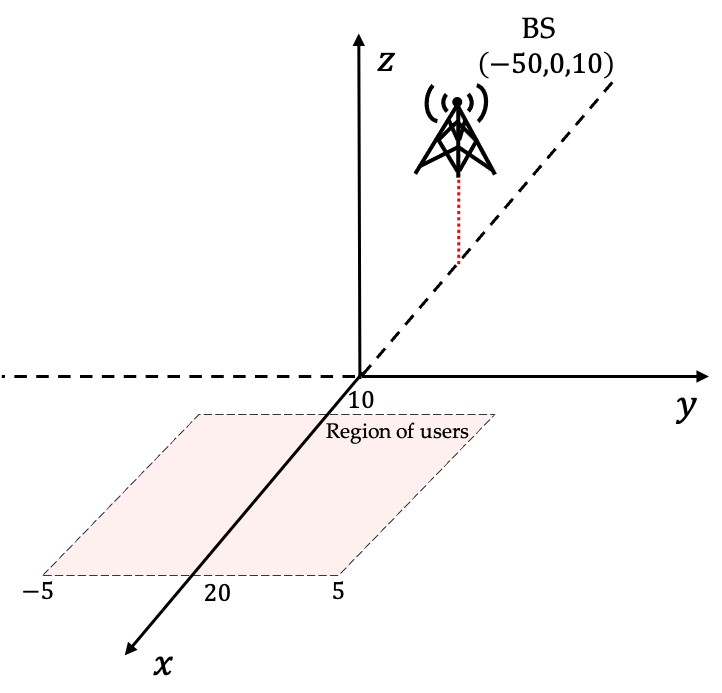}}
\caption{Participants Distribution in Simulation Experiment.}
\label{layout}
\end{figure}

We use a three-dimensional Cartesian coordinate system to illustrate the relative position between users and BS involved in the simulation experiment, as shown in the Fig. \ref{layout}.

BS is placed at $(-50,0,10)$. For the assignment of users, we consider a uniform distribution of geographical location and data size: All $K$ devices are uniformly allocated in a rectangular Region (i.e. Region $\triangleq \{\left(x,y,0\right):100\leq x \leq 150,-25\leq y \leq 25\}$), and each device is assigned the same number of training data samples.

The channel coefficients are given by the small-scale fading coefficients (following the standard independent and identically distributed (i.i.d.) Gaussian distribution) multiplied by the square root of the path loss. Specifically, the free-space path loss model is adopted for the device-BS direct channels as
\begin{equation}
    PL_{DB} = G_{BS}G_{D}\left(\frac{3\times10^8}{4\pi f_c d_{DB}}\right)^{PL},
\end{equation}
where $G_{BS}=5dBi$ is the antenna gain at the BS; $G_D=0dBi$ is the antenna gain at the devices; $f_c=915MHz$ is the carrier frequency; $PL=3.76$ is the path loss exponent; and $d_{DB}$ is the distance between the device and the BS.

It is worth mentioning that in the setting, although users have the same data sample distribution and are evenly distributed in the same area, this area is far away from BS, so their channel state may cause errors in the transmission process to a certain extent, thus affecting FL performance. In this case, it is necessary to measure the SER of the device and make the selection of participating devices based on SER. We used the convolutional neural network to train and test the MNIST and Fashion-MNIST dataset respectively. The cross entropy function is used as the loss function of network training.

\subsection{Effect of Modulation Order on FL Performance}

\begin{figure}
    \begin{minipage}{0.5\linewidth}
        \centering
        \includegraphics[width=1.0\linewidth]{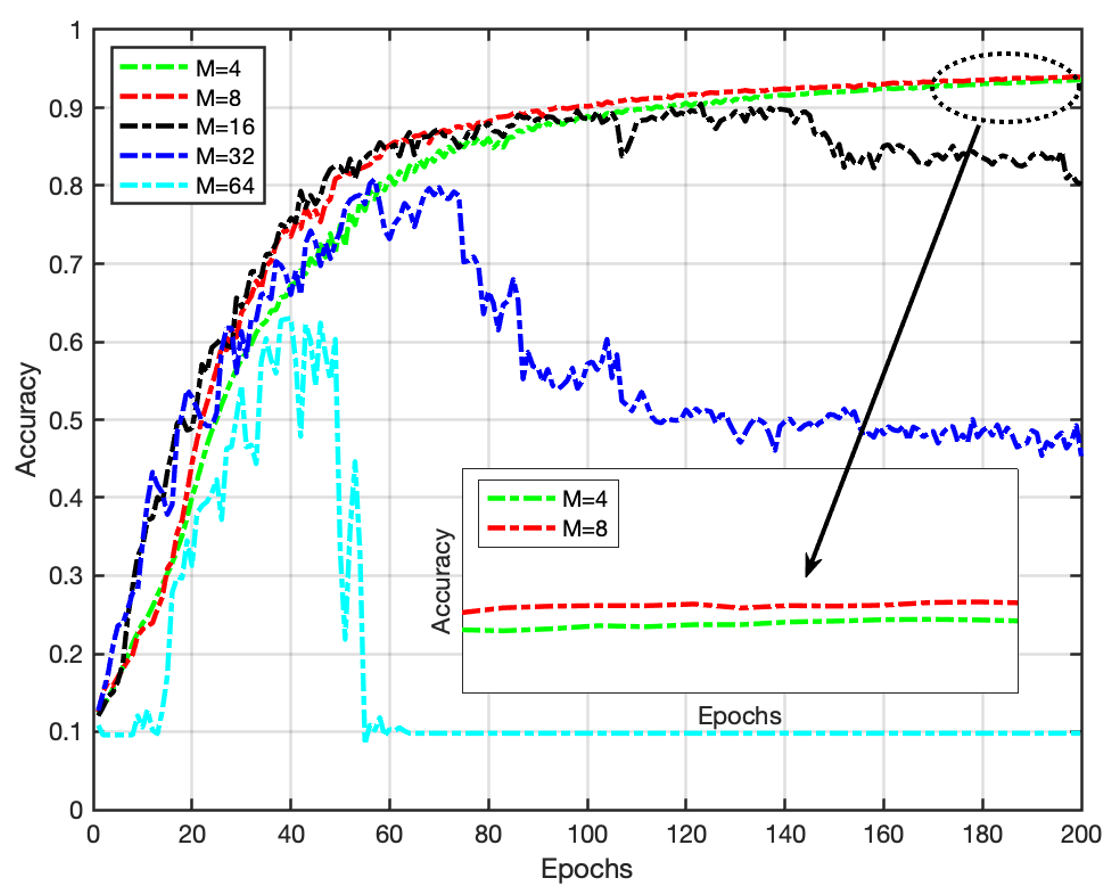}
        \subcaption{MNIST (q=21840)}
    \end{minipage}%
    \begin{minipage}{0.5\linewidth}
        \centering
        \includegraphics[width=1.0\linewidth]{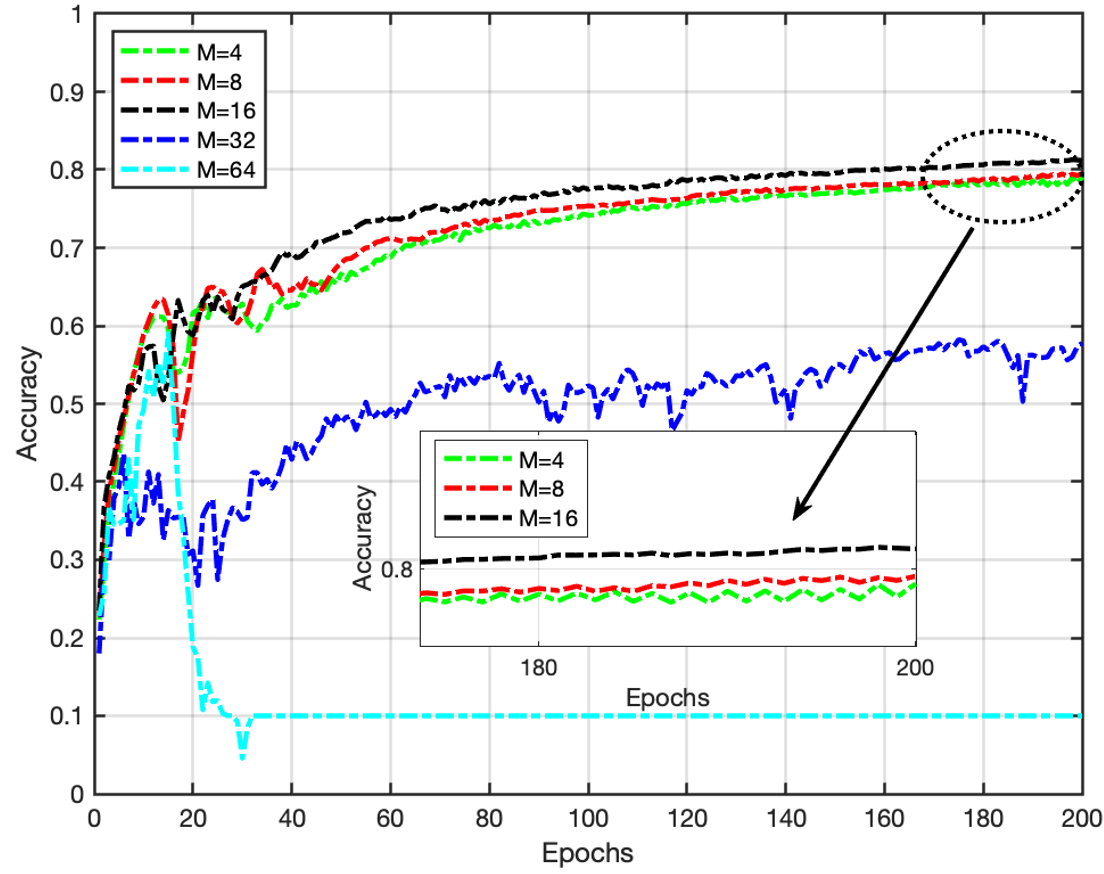}
        \subcaption{Fashion-MNIST (q=28938)}
    \end{minipage}
    \caption{The effect of the gradients on FL performance after transmission with different modulation orders.}
    \label{exp0}
\end{figure}

The experiment in this section aims to illustrate the effects of different modulation orders on FL performance in the case of multi-bit quantization. As mentioned in Section II, each floating gradient value to be transmitted will be quantized into a binary symbol consisting of $b_{signal}$ bits, then mapped to a QAM modulation signal of $M=2^{b_{mod}}$. If $b_{mod}\neq b_{signal}$ is used as the unit of transmission gradient with the M-ary modulation symbol, uncontrollable cumulative errors will occur. In order not to introduce additional errors due to the mismatch between the quantization level and the modulation order, we let $b_{mod}= b_{signal}$. For exploring the influence of different quantization levels on FL performance, we can directly observe the influence of different modulation orders on FL performance. The experimental results are shown in Figure \ref{exp0}.

Figure \ref{exp0}(a) shows the results of training and testing on the MNIST, and Figure \ref{exp0}(b) shows Fashion-MNIST. As can be seen from the figures, different modulation orders will have a direct impact on FL performance. For the MNIST, FL can obtain better accuracy and faster convergence rate when the gradient is modulated by 4-ary QAM and 8-ary QAM, and the larger modulation order will produce better performance. However, for higher order QAM modulation, such as $M=16$, $M=32$ and $M=64$, FL performance decreases significantly with increasing modulation order. The reason for this is that, with the increase of the number of $b_{signal}$ or $b_{mod}$, in addition to making the gradient symbol contain more information, it may bring more risk of symbol error accumulation due to code error, thus significantly increasing the symbol error rate $SER$, which is unfavorable to FL.

For the Fashion-MNIST, the difference is that the accuracy of FL is higher when $M=16$ than when $M=4$and $M=8$. Apart from this, FL performance shows the same trend with the modulation order as it does with MNIST. The reason for this is that, first of all, the parameter dimension of the learning task corresponding to the Fashion-MNIST is larger than that of MNIST. Secondly, for the same communication system (that is, the noise and interference conditions remain unchanged), when the number of gradient symbols to be transmitted increases, the importance of the information contained in it will also increase, that is, the more the number of symbols to be transmitted, the more information the symbols need to contain, which will offset the cumulative risk of symbol errors caused by code errors to a certain extent.

To sum up, for FL which quantizes gradient parameters in multiple bits, it is necessary to select the appropriate quantization level and modulation order under the same learning task, and the quantization level can be appropriately increased but not too high. When the dimension of the gradient parameter increases due to the difficulty of the learning task, a relatively higher quantization series can be appropriately chosen to quantify the gradient. 

\subsection{Device Selection Mechanism}

\begin{figure}
    \begin{minipage}{0.5\linewidth}
        \centering
        \includegraphics[width=1.0\linewidth]{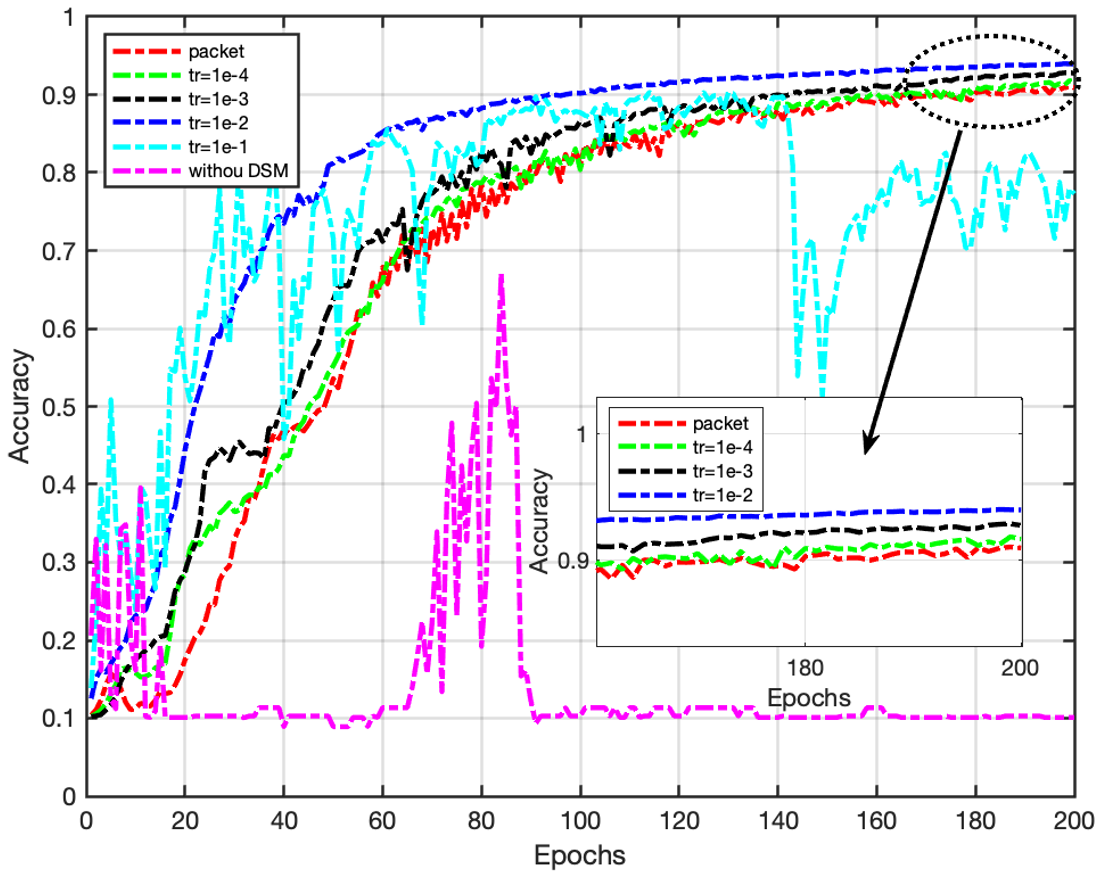}
        \subcaption{MNIST (q=21840)}
    \end{minipage}%
    \begin{minipage}{0.5\linewidth}
        \centering
        \includegraphics[width=1.0\linewidth]{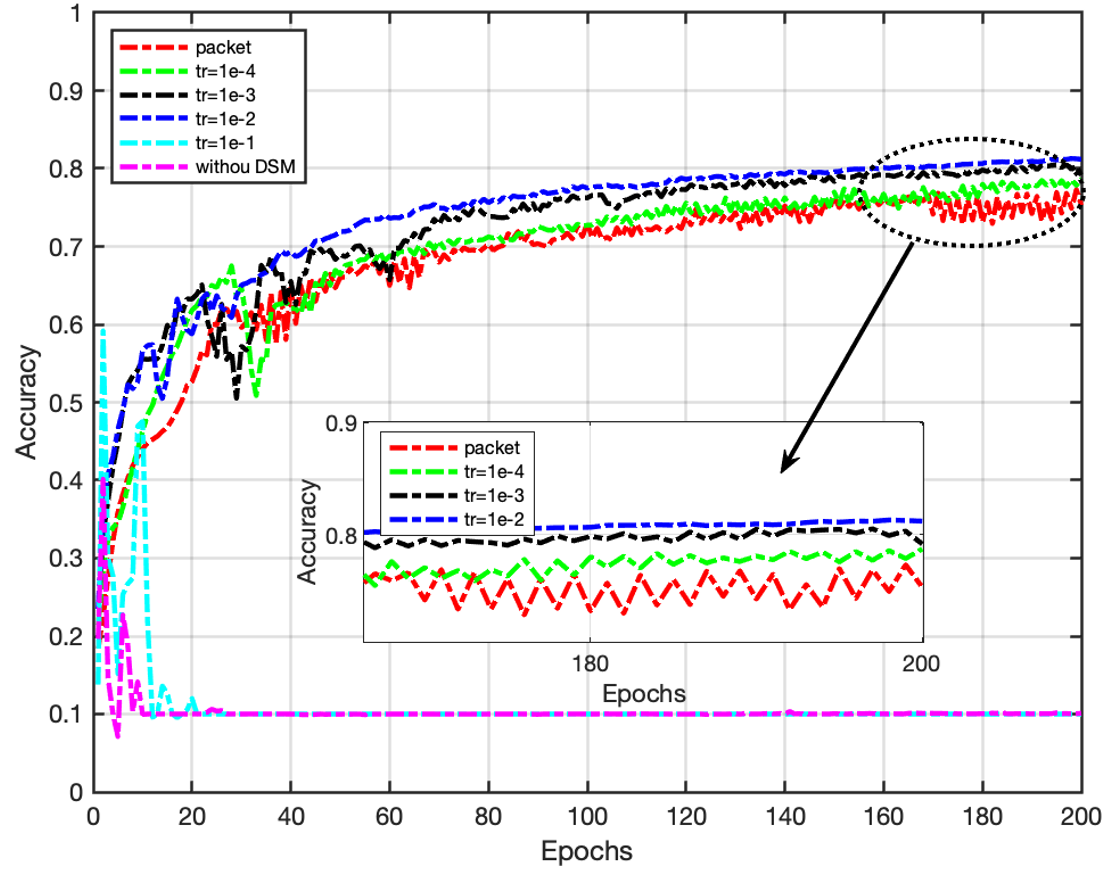}
        \subcaption{Fashion-MNIST (q=28938)}
    \end{minipage}
    \caption{The FL performance under the device selection mechanism proposed in this paper.}
    \label{exp1}
\end{figure}

In this section, the accuracy is used to illustrate the necessity and superiority of the proposed device selection mechanism training and testing on MNIST (Fig. \ref{exp1}(a)) and Fashion-MNIST (Fig. \ref{exp1}(b)) respectively. The experiment simulates the following benchmarks: 1) Simulates the device selection mechanism based on packet error rate in \cite{TWC} (red line); 2) No device selection mechanism, that is, all devices are involved in the FL update process (purple line); 3) Set schemes corresponding to 4 groups of different $tr_{SER}$ values (green, black, blue and cyan lines) according to the device selection mechanism proposed in this paper.

Firstly, for the convergence rate of FL, in the early stage in line with the more users involved in the faster the convergence rate, this is the general knowledge of the law, because the increase in the amount of data is conducive to the update of FL. However, both the $tr=1e-1$ (cyan line) benchmark and benchmark group 2 have more users participating in FL than the other benchmarks, but the convergence performance is far worse. This is because although the number of participating users increases, the communication conditions of these increased users are bad, resulting in more errors, which is not conducive to the update of FL. From the above analysis, it is not difficult to see that the performance of FL increases first and then decreases with the increase of the number of participating users, and reaches the best performance when $tr=1e-2$. Therefore, the device selection mechanism proposed in this paper is necessary in the case of wireless communication errors.

Secondly, allowing all users to participate in the FL (that is, without device selection) is the worst schedule, because it contains too many users with poor communication conditions, which will seriously affect the FL update. Meanwhile, the performance of FL after the SER-DSM is significantly better than that of base group 1. This is because proposed SER-DSM, while taking into account the need to reject users with poor transmission conditions, also takes into account the desire to accommodate as much data as possible for FL updates, so as to avoid data waste to the greatest extent. In other words, the device selection mechanism proposed in this paper is more inclusive. Finally, from the point of view of communication noise, we can find that when the noise power is constant, the more the number of users, the greater the oscillation amplitude of the FL convergence curve. In addition, after testing with the Fashion-MNIST dataset, the same change trend was also obtained. 

\section{Conclusions and Future Work}

This paper analyzes the advantages of multi-bit quantization of gradient and multi-QAM modulation in FL, and the effects of the number of quantization bits or modulation order on FL performance. On the basis of multi-bit quantization, we design an inclusive SER-DSM, and illustrate the effectiveness through convergence analysis. Finally, the experiments shows that: 1). When the gradient is quantized in multi-bit, the number of quantized bits needs to be determined according to the specific learning task; 2) Compared with the schedule of no device selection, the SER-DSM proposed in this paper has a certain necessity. Meanwhile, it results in better performance for FL than a device selection mechanism based on packet errors.

\bibliographystyle{IEEEtranTIE}
\bibliography{references}\ 

\vspace{12pt}

\end{document}